\documentclass[twocolumn]{aastex631}
\usepackage{amsmath}
\usepackage{multirow}
\usepackage{amssymb}
\usepackage{graphicx}
\usepackage{color}
\usepackage{float}
\usepackage{comment}

\definecolor{darkgreen}{rgb}{0,0.35,0}
\usepackage{CJK}

\begin{document}
\begin{CJK*}{UTF8}{gbsn}

\title{Spectral Appearance of Self-gravitating Disks Powered by Stellar Objects: Universal Effective Temperature in the Optical Continuum and Application to Little Red Dots}

\shorttitle{Self-gravitating, Star-forming Disks as LRDs}
\shortauthors{Chen et al.}

\author{Yi-Xian Chen} 
\affiliation{Department of Astrophysical Sciences, Princeton University,  4 Ivy Lane, Princeton, NJ 08544, USA}
\author{Hanpu Liu} 
\affiliation{Department of Astrophysical Sciences, Princeton University,  4 Ivy Lane, Princeton, NJ 08544, USA}
\author{Ruancun Li} 
\affiliation{Max-Planck-Institut f{\"u}r extraterrestrische Physik, Gie{\ss}enbachstra{\ss}e 1, 85748 Garching bei M{\"u}nchen, Germany}
\author{Bingjie Wang} 
\thanks{NHFP Hubble Fellow}
\affiliation{Department of Astrophysical Sciences, Princeton University,  4 Ivy Lane, Princeton, NJ 08544, USA}
\author{Yilun Ma}
\affiliation{Department of Astrophysical Sciences, Princeton University,  4 Ivy Lane, Princeton, NJ 08544, USA}
\author{Yan-Fei Jiang}
\affiliation{Center for Computational Astrophysics, Flatiron Institute, New York, NY 10010, USA}
\author{Jenny E. Greene}
\affiliation{Department of Astrophysical Sciences, Princeton University,  4 Ivy Lane, Princeton, NJ 08544, USA}
\author{Eliot Quataert}
\affiliation{Department of Astrophysical Sciences, Princeton University,  4 Ivy Lane, Princeton, NJ 08544, USA}
\author{Jeremy Goodman}
\affiliation{Department of Astrophysical Sciences, Princeton University,  4 Ivy Lane, Princeton, NJ 08544, USA}
\begin{abstract}
{We revisit the spectral appearance of extended self-gravitating accretion disks surrounding compact central objects such as supermassive black holes.} 
Using dust-poor opacities, 
we show that all optically thick disk solutions possess a universal outer effective temperature of $T_{\rm eff}\sim 4000-4500$K, closely resembling compact, high-redshift sources known as Little Red Dots (LRDs). 
Assuming the extended disk is primarily heated by stellar sources, this ``disk Hayashi limit" 
fixes the dominant optical continuum temperature of the disk spectrum independent of accretion rate $\dot{M}$, 
central mass $M_\bullet$, and disk viscosity $\alpha$, and removes the parameter-tuning required in previous disk interpretations of LRDs. 
{The formation and accretion of embedded stellar objects can both power the emission of the outer disk and hollow out the inner disk, 
suppressing variable UV/X-ray associated with a standard quasar. }
The resulting disk emission is dominated by a luminous optical continuum 
while a separate, non-variable UV component arises from stellar populations on the nuclear to galaxy scale.
We map the optimal region of parameter space for such systems
and show that LRD-like appearances naturally emerge for $\dot{M}/\alpha \gtrsim 0.1 M_\odot /{\rm yr}$, a threshold insensitive to $M_\bullet$, 
below which the system may transition into classical non-self-gravitating AGN disks, potentially a later evolution stage. 
We expect this transition to be accompanied by the enhancement of metallicity 
and production of dust, giving rise to far infrared emission. 
This picture offers a physically motivated and quantitative framework connecting LRDs with AGNs and their associated nuclear stellar population.
\end{abstract}

\section{Introduction}
\label{sec:intro}

Since its launch, the James Webb Space Telescope (JWST) has 
uncovered a previously unrecognized class of compact, red sources known as ``Little Red Dots” (LRDs; \citealt{Matthee2024}), which are now being identified in large numbers. These objects are unresolved in rest-optical images \citep[e.g.,][]{Kokorev2024, Akins2025,Hviding2025,deGraaff2025} and show broad Balmer emission lines \citep[e.g.,][]{Greene2024, Lin2024aspire} together with a distinctive V-shaped continuum: they are blue in the rest-frame ultraviolet (UV) but extremely red in the rest-frame optical \citep{Labbe2023, Kocevski2024, Hviding2025}, with the red optical continuum dominating the total spectral output \citep{Greene2025}. Their extreme compactness and unusual spectral energy distributions have attracted considerable attention, motivating a range of new physical scenarios for their nature and evolution.

At first glance, 
the most straightforward interpretation is a conventional dust-reddened AGN accretion disk plus a stellar population component from the host galaxy \citep[e.g.][]{Wang2024:ub, Labbe2024, Ma2025lrd404, WangdeGraaff2025}. 
However, the substantial dust extinction implied by these models is difficult to reconcile with the stringent constraints on dust emission in the rest-mid and far-infrared (FIR) from deep JWST/MIRI and ALMA observations \citep{Williams2024,Akins2025,Casey2025,Setton2025,WangdeGraaff2025,Xiao2025}.
This motivates scenarios in which the optical/red bump arises from thermal emission produced by dense, gas-dominated envelopes in an almost dust-free environment. 
Such configurations resemble 
scaled-up versions of stellar envelopes at their Hayashi limits with a characteristic $T_{\rm eff} \sim 5000$K regulated by H$^-$ opacity, 
and have been referred to as ``black hole stars" or ``quasi-stellar''-type models \citep{deGraaff2025, Naidu2025, Begelman2025, Kido2025, Liu2025, Nandal2025, WangJM2025}.
This class of models is also motivated by strong Balmer breaks observed in a subset of LRDs \citep{Inayoshi2025a,Naidu2025, Ji2025, Cliff}, 
a feature that poses a serious challenge for both standard AGN continua and conventional post-starburst stellar populations.
On the blue side, 
the lack of strong rest-UV continuum variability in most systems \citep{Furtak2025, Ji2025, Zhang2025a,Zhang2025b} 
adds further tension with a conventional AGN picture and provides important additional constraints on viable physical models.

An alternative to the spherical scenario is to invoke an optically thick accretion disk \citep{Inayoshi2025CBD,Liu2025} as an interpretation for LRDs, in which the spectral energy distribution (SED) is obtained by integrating the thermal emission from individual disk annuli over a prescribed radial profile.
The difficulty with this scenario is that, 
for a standard disk powered by accretion, $T_{\rm eff} \propto R^{-3/4}$ \citep{SS1973}, 
and the spectral shape of the disk will be governed by the effective temperature at the inner boundary $R_{\rm in}$. 
Reproducing the observed red optical continuum therefore requires imposing an inner truncation radius such that $T_{\rm eff}\sim 5000$K, 
an assumption that lacks a clear physical justification from first principles. 

The ``inner cutoff" issue may be alleviated for extended 
self-gravitating disks outside the standard AGN disk's self-gravitating radius $R_{\rm sg}$, 
beyond which embedded stellar-objects formed from gravitational instability (GI) may replace accretion to
power the disk emission and regulate the Toomre parameter $Q(R> R_{\rm sg})\sim 1$ \citep{Sirko2003,GoodmanTan2004, Thompson2005,Gilbaum2022,ChenLin2024,Zhou2024, Epstein-Martin2024}. 
For these disk regions, 
$T_{\rm eff}$ profiles can be flatter than $R^{-1/2}$ and even invert, 
causing the SED peak to be determined by the effective temperature at the 
\textit{outer cutoff} $R_{\rm out}$. 

This may appear to generate another issue of fine-tuning: if $R_{\rm out}$ were allowed to extend to arbitrarily large radii, the corresponding ``dominant" $T_{\rm eff}$ 
could decline accordingly, shifting the SED peak into the far-infrared and making the spectrum too red to match those of LRDs. Indeed, 
\citet{Zhang2025} 
apply a constant accretion rate, self-gravitating disk model from \citet{Sirko2003} to fit LRD spectra and find that $T_{\rm eff} (R_{\rm out}) \sim 4500$K 
constrains $R_{\rm out}$ to be no more than a factor of a few larger than $R_{\rm sg}$, which also lacks a physical explanation.

In this work,
we argue that $T_{\rm eff}(R_{\rm out})$ is not freely adjustable. 
Once $R_{\rm out}$ is defined as the outer edge of the optically thick portion of the disk capable of producing thermalized emission, 
the effective temperature there is set by the transition to an optically thin regime. 
Beyond this radius, the emission from the disk will resemble stellar populations. 
As long as the opacity is dust-free, 
consistent with the observed lack of FIR emission, the resulting
$T_{\rm eff}$ dominating the thermalized emission is constrained to a nearly universal value.
This universality intrinsically arises from the characteristic shape of the $\rm H^-$
opacity and can therefore be viewed as the disk analog of the Hayashi limit, 
directly paralleling the mechanism generating universal $T_{\rm eff}$
in spherical envelope models of LRDs. 

{More generally, this phenomenology 
do not strictly require a central SMBH and may apply to 
self-gravitating nuclear disks hosted by other central potentials, 
such as dense stellar clusters or supermassive stars \citep{Zwick2025}. 
Nevertheless, we note that}
the AGN interpretation is especially motivated by the observed evolutionary connection between LRDs and the high-redshift quasar population\citep{Ma2025,Umeda2025}, 
which implies that LRDs are likely an early phase of SMBH growth directly followed by standard AGN-like appearances after $z\sim 4$ \citep{Inayoshi2025abundance}.
However, relevance to standard AGNs is a double-edged sword: 
another difficulty in applying the \citet{Sirko2003} disk model
to LRDs is precisely that it connects to a standard AGN disk inside $R_{\rm sg} \sim 1000 R_g$. 
\citet{Zhang2025} relies on this part of disk emission to explain the UV component of the LRDs, 
but the standard AGN disk picture may be inconsistent with the 
lack of variability in the UV \citep[e.g.][]{Burke2025}. 

\begin{figure*}
    \centering
 \includegraphics[width=1\textwidth]{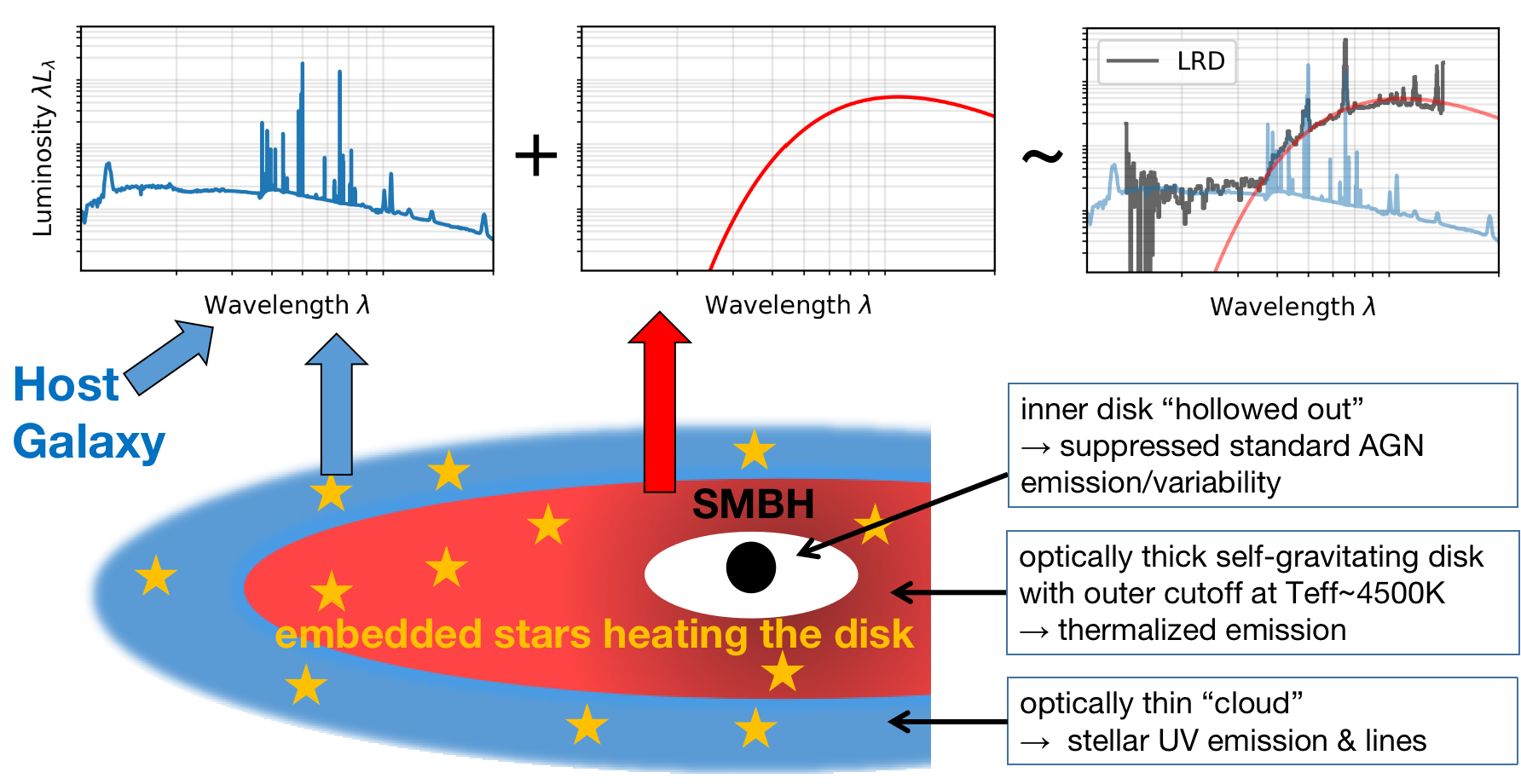}
    \caption{Schematic illustration of our proposed physical interpretation of LRDs. 
    The dominant red/optical emission arises from an optically thick, 
    self-gravitating disk with heating primarily supplied by embedded stellar populations (see \S \ref{sec:outer_disk}). 
    {The inner disk is strongly depleted by star formation, 
    suppressing the classical variable UV/X-ray emission from a 
    standard AGN below current observational constraints, although weak residual accretion onto the central source is still allowed (see \S \ref{sec:inner_disk}).}. 
    A separate UV component can originate from stellar populations in a surrounding optically thin, diffuse cloud that connects the disk to the nuclear stellar population.
    The top row shows a schematic spectral decomposition for illustration purposes. 
    We use the observed spectrum of a representative LRD source RUBIES-40579 \citep{WangdeGraaff2025} (black lines, top right panel) to illustrate how the total emission may be qualitatively understood as the sum of a thermalized red/optical disk component and a stellar-dominated UV component. 
    This decomposition is not intended as a formal spectral fit,  
    only a visual guide to the proposed physical picture. 
    { Although we illustrate the system with a central SMBH, 
    the model also apply to other compact central potentials, 
    such as supermassive stars or dense nuclear stellar clusters. 
    Nevertheless, throughout most of this work we focus on the SMBH case 
    because it connects more directly, in an evolutionary sense, to the standard quasar population.}} 
    \label{fig:schematics}
\end{figure*}

We propose that a more self-consistent picture is one in which the outer $Q\sim 1$ disk produces the optical/red bump, while the inner standard accretion disk is ``hollowed out" (see schematic in Figure \ref{fig:schematics}). Revisiting the long-standing idea that star formation in the AGN disk can deplete mass inflow, 
we argue that this process can naturally suppresses variable X-ray and UV radiation from the inner disk. 
In our picture, 
the observed UV emission instead arises from stellar populations in the extended optically thin cloud beyond the optically thick self-gravitating disk, 
which connects with nuclear stellar environments and emits stellar radiation that is \textit{not} reprocessed into a thermal continuum. 
The boundary between these regions has a universal $T_{\rm eff}\sim 4500$K, while additional
emission lines may arise from surrounding optically thin material directly outside the optically thick region. 
Together, 
these elements provide a coherent explanation for the distinctive optical properties of LRDs. 
While \citet{Thompson2005} first presented disk models with $\dot{M}$ being depleted inwards, 
their application was aimed at ultraluminous infrared galaxies (ULIRGs) and employed dusty opacities, 
allowing $R_{\rm out}$ to extend to large radii and thereby producing much FIR emission. 
In contrast, dust-free opacity prevents $R_{\rm out}$ from
drifting to large scales and avoids the FIR-bright regime, 
enabling star-forming disk models to remain consistent with the appearance of LRDs. 


This letter is organized as follows: 
In \S \ref{sec:universal_Teff} we sketch out a proof that dust-free opacity can determine a universal $T_{\rm eff}(R_{\rm out})$ for optically thick self-gravitating disk models, insensitive to all other assumptions. 
{In \S \ref{sec:outer_disk} and \S \ref{sec:inner_disk}, we demonstrate that the continuum emission of this system
is consistent with LRDs over a plausible range of parameters, }
and propose an evolutionary scenario in which LRDs naturally transition into standard AGNs. 
Finally, in \S \ref{sec:discussion}, we summarize our general picture and 
discuss future prospects in this framework for 
(i) modeling partially thermalized irradiation from stellar populations relevant to UV emission and spectral lines, 
(ii) effect of metallicity and dust formation in the emergence of FIR emission and deviation from LRD-like appearances, (iii) implications for cosmological abundances and iv) connections to the topic of stellar-object evolution in quasar disks.

\section{Universal effective temperature for the boundaries of optically thick disk regions}
\label{sec:universal_Teff}

In this section, we will use a local model based on a given midplane density 
$\rho$ and temperature $T$ 
to illustrate the universal outer $T_{\rm eff}$ phenomenon, 
regardless of the details of the radial distribution $T(R), \rho(R)$. 

Using the approximate condition for marginal gravitational instability ($Q\approx 1$; \citealt{Goodman2003})

\begin{equation}
    \Omega^2 = 2\pi G \rho
    \label{eqn:self-gravity}
\end{equation}

and the Eddington Equation of state

\begin{equation}
    \rho c_s^2 = \dfrac{\rho \mathcal{R} T}{\mu} + \dfrac{aT^4}{3},
\end{equation}

we can write down the vertical scale height $H = c_s/\Omega$ in local parameters

\begin{equation}
    H = \sqrt{ \dfrac{a T^4}{6\pi G\rho^2}+\dfrac{\mathcal{R} T}{2\pi \mu G \rho}}
    \label{eqn:scale_height},
\end{equation}

where $\mathcal{R}$ is the gas constant and $\mu$ is the molecular weight, taken to be 0.6 in subsequent numerical calculations. For an optically thick region, 
photon diffusion leads to $ T^4 \approx \tau T_{\rm eff}^4\approx \kappa \rho H T_{\rm eff}^4$, where $\sigma T_{\rm eff}^4$ is the flux from the photosphere of the disk. 

If we take representative Rosseland mean opacity $\kappa_R(\rho, T)$ in the expression for optical depth, 
the effective temperatures can be expressed as a function of $\rho, T$

\begin{equation}
T_{\mathrm{eff}}=\left[\frac{T^4}{\kappa_R(\rho, T) \rho \sqrt{\frac{a T^4}{6 \pi G \rho^2}+\frac{\mathcal{R} T}{2 \pi \mu G \rho}}}\right]^{1 / 4},
\label{eqn:Teff_T_rho}
\end{equation}

which is self-consistent as long as $T_{\mathrm{eff}}< T$, else the region is implied to be optically thin and the 
diffusion approximation 
will no longer be valid. 
This can be simplified to be

\begin{equation}
   T_{\mathrm{eff}} = \left[\dfrac{T^2}{\kappa_R(\rho, T) \sqrt{\frac{a}{6 \pi G }}}\right]^{1 / 4}
   \label{eqn:Teff_T_rho_rad}
\end{equation}

in the radiation-pressure-dominated regime and

\begin{equation}
   T_{\mathrm{eff}} = T \left[{\dfrac{2\pi \mu \rho }{\kappa_R^2(\rho, T) \mathcal{R} T}}\right]^{1/8}
    \label{eqn:Teff_T_rho_gas}
\end{equation}

in the gas-pressure-dominated regime. Both will sharply decrease with temperature when $\partial \ln \kappa/\partial \ln T \gg 1$.

\begin{figure}
    \centering
 \includegraphics[width=0.47\textwidth]{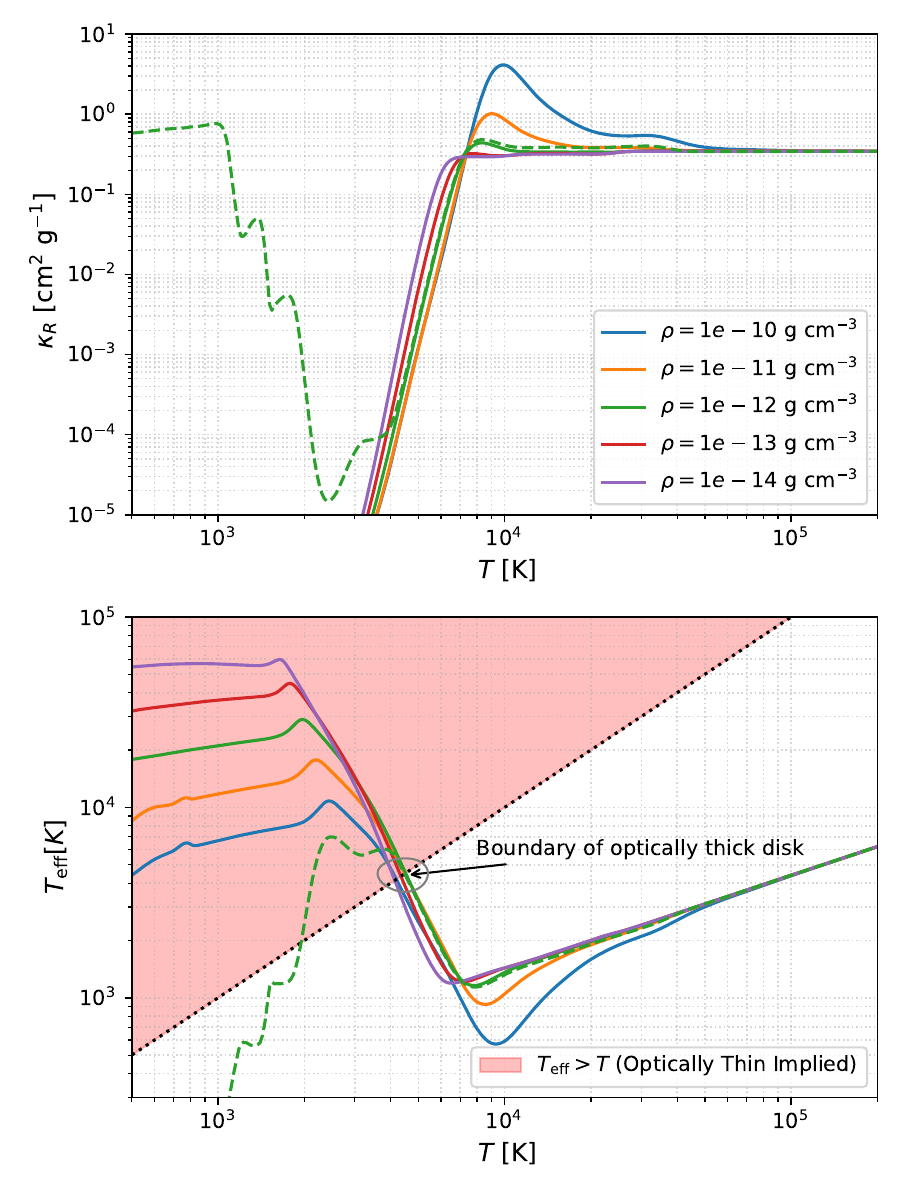}
    \caption{Top panel: solid lines indicate metal-free ($Z=0$) Rosseland mean opacities $\kappa_R(T)$ for different densities $\rho$. 
    A representative $\kappa_R(T)$ profile for solar metallicity opacity with dust for $\rho =10^{-12}~$g~cm$^{-3}$ is shown as green dashed line for comparison. 
    Lower panel: effective temperature calculated by Equation \ref{eqn:Teff_T_rho} for different densities. 
    The shaded region represents solutions with $T_{\rm eff} > T$ that are no longer consistent with the optically thick assumption, implying a stellar-UV rather than thermalized emission. 
    For $Z=0$ opacities, 
    there is a unique and universal transition at $T_{\rm eff}\approx T\approx 4000-5000$K (gray circle), regardless of density. } 
    \label{fig:Teff_thick}
\end{figure}

In Figure \ref{fig:Teff_thick}, 
we plot $\kappa_R(T)$ (top panel) as well as solutions of Equation \ref{eqn:Teff_T_rho} 
(lower panel) as functions of $T$ for different values of $\rho$. 
The fiducial opacity tables (solid lines) assumes zero metallicity, 
such that no dust grains form at low temperatures. 
As a result,
$\kappa_R(T)$ drops sharply and becomes extremely low at $ T < 2000$K. 
For comparison, 
we also show a representative opacity and effective-temperature solution at $\rho = 10^{-12}$g~cm$^{-3}$ using \textit{solar-metallicity} opacity that include dust grains, 
plotted as green dashed curves. 
The opacity at different metallicities are taken from the MESA database \citep{paxton_2024_13353788}, which is a compilation of tables from \citet{OPAL1993,OPAL1996,Ferguson2005}. 

The key universal feature for $T_{\rm eff}$ 
calculated from our fiducial opacity law is that regardless of $\rho$, 
most of the parameter space is consistently optically thick 
whereas all transition 
into the optically thin region must occur within the narrow parameter space where $T_{\rm eff}\sim T\sim 4000-4500$K due to the steep opacity law, 
suggesting a 
universal outer boundary $T_{\rm eff}$ for optically thick solutions, 
insensitive to the details of $\rho(R), T(R)$ and how mass flow is radially transported. 
This can be qualitatively taken as a self-gravitating disk version of the Hayashi limit. 
For high metallicity, 
dusty opacities (e.g. the green dashed lines), 
the opacity increases back up at low $T$ 
and allows for an outer optically thick branch of solutions that give rise to far infrared emission at $T_{\rm eff} < 2000$K. The dusty branch of solutions can extend to arbitrarily low midplane temperature
applicable to ULIRGs \citep{Thompson2005}, but is non-existent 
as long as dust-free opacities is applied.

We adopt the 
metal-free opacity (corresponding to the dust-free limit in a \textit{narrow} sense) for illustrative clarity in \S \ref{sec:universal_Teff}. 
Nevertheless, 
the emergence of a universal outer effective temperature is not restricted to low metallicity, 
but also persists in regimes where the gas remains 
\textit{broadly} dust-poor despite metal enrichment \citep{LeeChiangOrmel2014}. This is possible if metals fail to condense into grains at low temperatures and instead remain predominantly in molecular form. 
The development of a 
possible FIR-emitting dusty branch 
at non-negligible grain abundance is highly relevant to connecting our LRD framework to classical AGNs, 
and we return to this issue in \S \ref{sec:metal_dust}.

\section{Emission From the Outer Disk and General Constraints}
\label{sec:outer_disk}

For a global disk, the location of 
$R_{\rm out}$ can be determined by
three main parameters: 
the central mass $M_\bullet$, accretion rate at the outer boundary $\dot{M}(R_{\rm out})$, 
and the disk viscosity parameter $\alpha$. 
Following the argument in \S \ref{sec:universal_Teff}, 
we
define $R_{\rm out}$ as the point where $T = T_{\rm eff}$ for given $\dot{M}(R_{\rm out})$, and
we invoke a mass accretion prescription at $Q=1$:

\begin{equation}
    \dot{M} = \alpha (H/R)^3 M_\bullet \Omega = \alpha (H/R)^3 M_\bullet\sqrt{GM_\bullet/R^3}
    \label{eqn:local_mdot}
\end{equation}

which can be inverted to obtain 
$H(R; \dot{M}, M_\bullet, \alpha)$. 
We first assume $\Omega$ to be 
dominated by the potential of the point source $M_\bullet$, 
but will discuss later that this might not be the most general case.
Next, we rewrite Equation \ref{eqn:scale_height} as $H(R; M_\bullet, \alpha,T(R_{\rm out}))$ 
by imposing the self-gravitating condition (Equation \ref{eqn:self-gravity}). 
Combining this with Equation \ref{eqn:local_mdot} yields 
$R_{\rm out}(\dot{M}, M_\bullet, \alpha, T(R_{\rm out}))$, 
and $R_{\rm out}(\dot{M}, M_\bullet, \alpha)$ can be solved iteratively by adopting searching for the value of
$T(R_{\rm out})\approx 4000$K that satisfy $T = T_{\rm eff}(T)$, 
making use of Equation \ref{eqn:Teff_T_rho}.

{Given boundary conditions specified by $\dot{M}$, $M_\bullet$, $\alpha$, and $R_{\rm out}(\dot{M}, M_\bullet, \alpha)$, 
one can integrate a self-gravitating disk structure by introducing an additional continuity equation that prescribes $d\dot{M}/dR$ in terms of the local variables.
The constant-$\dot{M}$ case corresponds to $d\dot{M}/dR = 0$ \citep{Sirko2003}. 
In \citet{Thompson2005, ChenLin2024}, this assumption is instead replaced by a constant mass-to-light conversion efficiency $\epsilon$ for the stellar component, 
which provides heating to balance radiative cooling in the disk. 
Alternatively, one may impose a stellar number density distribution regulated by dynamical interactions \citep{Levin2003}.
Despite diverse possibility of this closure relation,
there exist generic scaling relations for the thermal emission from the outer star-forming disk large independent of these the detailed treatment. 
The key point is that 
the SED is dominated by annuli near the outer boundary $R_{\rm out}$ and, 
unlike the inner disk (see \S \ref{sec:inner_disk}), 
is relatively insensitive to the detailed radial profile of $\dot{M}(R)$}:

\begin{equation}
    L_{\rm disk} \approx 2 \pi f R_{\rm out}^2 \sigma T_{\rm eff}^4(R_{\rm out}),
\end{equation}

where $f<1$ takes into account the effective width of the annuli radiating at 
$T_{\rm eff}(R_{\rm out})$, 
$\Delta R\sim f R_{\rm out}$, that dominates the total luminosity of the disk.  
In the radiation pressure dominated scenario,  $R_{\rm out}$ can be expressed as 

\begin{equation}
R_{\rm out}=\left[\frac{G M_{\bullet}^{3 / 2}\dot{M}(R_{\rm out})}{\alpha}\left(\frac{3}{2 \pi a T^4(R_{\rm out})}\right)^{3 / 2} \right]^{2 / 9}.
\end{equation}

For universal $T(R_{\rm out})\approx T_{\rm eff}(R_{\rm out})\approx 4000$K, we have
(hereon we use $\dot{M}$ to represent the accretion rate at the outer boundary)

\begin{equation}
    R_{\rm out} \approx 0.03 {\rm pc} \left(\frac{\dot{M} / \alpha}{1 M_{\odot} \mathrm{yr}^{-1}}\right)^{2 / 9}\left(\frac{M_{\bullet}}{10^6 M_{\odot}}\right)^{1 / 3}\left(\dfrac{T_{\rm eff}}{4000{\rm K}}\right)^{-3/4},
    \label{eqn:Rout}
\end{equation}
and correspondingly,  

\begin{equation}
\begin{aligned}
    L_{\rm disk} &\approx  2\times 10^{44} \mathrm{erg} \mathrm{s}^{-1} \left(\frac{f}{0.2}\right) \left(\frac{\dot{M} / \alpha}{1 M_{\odot} \mathrm{yr}^{-1}}\right)^{4 / 9}\\&\times \left(\frac{M_{\bullet}}{10^6 M_{\odot}}\right)^{2 / 3}\left(\frac{T_{\text {eff }}}{4000 \mathrm{~K}}\right)^{5 / 2}\\
    &= 2L_{\rm Edd} (M_\bullet) \left(\frac{f}{0.2}\right)\left(\frac{\dot{M} / \alpha}{1 M_{\odot} \mathrm{yr}^{-1}}\right)^{4 / 9}\\&\times \left(\frac{M_{\bullet}}{10^6 M_{\odot}}\right)^{-1 / 3}\left(\frac{T_{\text {eff }}}{4000 \mathrm{~K}}\right)^{5 / 2}.
    \end{aligned}
    \label{eqn:Ldiskanalytical}
\end{equation}

With this estimate, we can put very general constraints on the viable parameter space for self-consistency. 
When $\dot{M}/\alpha$ is too large, 
$L_{\rm disk}$ becomes significantly higher than the Eddington luminosity of the SMBH. 
This has the consequence of requiring a total luminous mass $>M_{\bullet}$ (e.g. massive stars, stellar-mass BHs at their Eddington luminosity) 
in the outer disk as heat source to support $L_{\rm disk}$. 
The radiation field of embedded stars may be anisotropic to allow for super-Eddington luminosity in polar solid angles \citep{ChenJiangGoodman2025}, 
but this can only 
relax the threshold by order of a few. 
Eventually, as $L_{\rm disk}\gg L_{\rm Edd}$, 
assuming $\Omega$ to be dominated solely by the SMBH's gravitational potential will no longer be self-consistent. 
Thus, 
the current framework is fully applicable only for moderate values of $\dot{M}/\alpha$. 
This limitation is not a physical prohibition; 
rather, it implies that modeling systems with very large $\dot{M}/\alpha$ requires a more complete and dynamical treatment 
that self-consistently follows the transition from massive star-dominated to a SMBH-dominated potential with a more general $\Omega(r)$ profile, 
which we will address in future works.
As we will elaborate later in this section, such a regime likely corresponds to the earliest phases of LRD evolution,  
where $\dot{M}$ 
is high and starburst is rapid. 
As the accretion rate declines and/or angular-momentum transport becomes more efficient, 
the total stellar mass may be self-regulated towards moderate levels. 
We take this transition limit for $L_{\rm disk} < 3 L_{\rm Edd}$ as 

\begin{equation}
     {\dot{M}}/{\alpha} < 5 M_{\odot} \mathrm{yr}^{-1}  \left(\frac{T_{\text {eff }}}{4000 \mathrm{~K}}\right)^{- 45 / 8} \left(\frac{M_\bullet}{10^6 M_\odot}\right)^{3 / 4} \left(\dfrac f{0.2}\right)^{-9/4} \,.
      \label{eqn:Maupperlimit}
\end{equation}


Another physical limit on $\dot{M}/\alpha$ arises from the requirement 
that a substantial portion of the disk must remain self-gravitating. 
This is guaranteed for radiation pressure dominated disks: 
combining Equation \ref{eqn:Rout} with the viscous support condition 
(see Equation \ref{eqn:viscous_criterion}), 
a radiation pressure dominated disk would not be viscously supported 
at $R_{\rm out}$ for any accretion rates lower than $\sim c^3/G$, 
an unphysically enormous rate.
Practically, the disk becomes marginally viscously supported close to $R_{\rm out}$ only if
gas pressure were to dominate immediately within $R_{\rm out}$, 
a premise that 
can be avoided as long as 

\begin{equation}
    \rho(R_{\rm out}) = \dfrac{M_\bullet}{2\pi R_{\rm out}^3}< 1.2\times 10^{-12} {\rm g/cm}^3 \left(\dfrac{T_{\rm eff}}{4000{\rm K}}\right)^3
\end{equation}

such that the material is radiation dominated at $T\approx T_{\rm eff} \approx 4000$K. 
This requires

\begin{equation}
    {\dot{M}}/{\alpha} > 0.1 M_{\odot} \mathrm{yr}^{-1} \left(\frac{T_{\text {eff }}}{4000 \mathrm{~K}}\right)^{- 9 / 8} 
    \label{eqn:Malowerlimit}
\end{equation}

Below this threshold relatively independent of the SMBH mass, 
the disk may behave as a classical AGN, possibly representing the late evolutionary stages of LRDs \citep{Fu2025}. 

\begin{figure}
    \centering
 \includegraphics[width=0.47\textwidth]{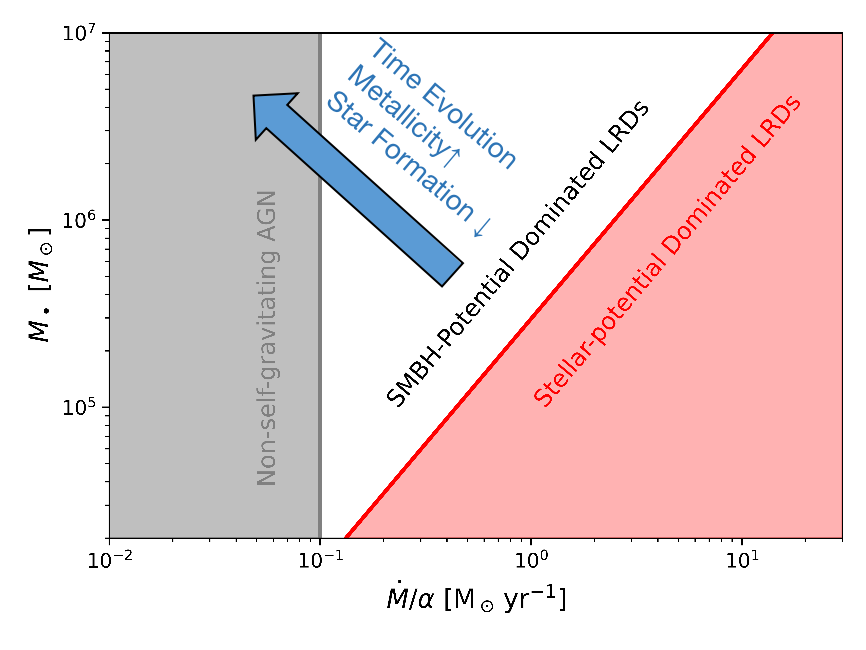}
    \caption{Parameter space of SMBH accretion disk in the $(\dot{M}/\alpha,M_\bullet)$ plane.
    Red indicates a regime of dynamical starburst where luminous mass $> M_\bullet$ is needed to support $L_{\rm disk}$ (Equation \ref{eqn:Maupperlimit}), 
    while the gray region marks the transition to a non–self-gravitating AGN disk (Equation \ref{eqn:Malowerlimit}). 
    These regimes may reflect sequential stages of SMBH disk evolution, transitioning from LRD-like systems to standard AGN disks.} 
    \label{fig:parameter}
\end{figure}

We now more generally illustrate the constraints from Equations \ref{eqn:Maupperlimit} and \ref{eqn:Malowerlimit} 
in Figure \ref{fig:parameter}. 
The white region delineates the parameter space in which our results most robustly apply, 
namely LRD systems with gravity dominated by the SMBH potential. 
Within the red region, sustaining 
$L_{\rm disk}$ would require the mass of luminous stellar sources to exceed $M_\bullet$. 
The dynamical structure of such systems would instead be governed by the stellar potential, 
leading to a flatter angular-frequency profile $\Omega(R)$ profile than the Keplerian case. 
Nevertheless, 
the argument in \S \ref{sec:universal_Teff} does not hinge on particular rotation profiles so we still expect LRD-like optical continuum properties. 

At the opposite extreme, 
the lower bound on $\dot{M}/\alpha$ is a limit beyond which the system transitions into a standard, 
non–self-gravitating AGN disk (gray region). 
The white region narrows toward lower $M_\bullet$, 
with the two boundaries approaching each other for $M_\bullet \lesssim 10^4 M_\odot$. 

Taken together, 
the three regimes likely represent successive phases of the coupled LRD-AGN evolution. 
We envisage that a finite gas reservoir starts to fuel a SMBH from large radii. 
During the earliest stage, 
the disk experiences strong external infall and accretion flows are converted into rapid starbursts in the outer regions. 
{As the stellar mass builds up and becomes comparable to or exceeds the SMBH mass (red region), global non-axisymmetric instabilities, such as stellar bars and spirals, 
are expected to develop and enhance angular momentum transport 
(which can be qualitatively understood as enhacing  effective $\alpha$, but also see Equation \ref{eqn:globalvr}) \citep{GoodmanRafikov2001, HopkinsQuataert2010} 
\footnote{\citet{Rozyczka1995} also showed that supernova occurring in quasar disks 
can enhance angular momentum transport.}. 
As a result, the system evolves towards lower $\dot{M}/\alpha$, 
while simultaneously feeding and growing the SMBH through both gas accretion and the capture of stellar material, 
thereby transitioning into the regime where SMBH potential dominates (white region) \footnote{We note that a central SMBH is not strictly required for the initial \textit{formation} 
of such systems. 
In fact, a
self-gravitating stellar disk may instead provide a viable pathway for seeding the SMBH. 
{Nevertheless, 
as discussed above, 
enhanced angular momentum transport will drive rapid inflow and efficiently feed any alternative central object (compact stellar clusters or supermassive stars
that can later collapse into a SMBH), 
leading the system toward the parameter space more directly relevant to AGNs.}}.}

{If very high transport efficiency were to be sustained, the system could in principle transition rapidly into the gray region. 
However, 
we expect the rapid inflow associated with dynamical starbursts to be self-limiting. 
As the SMBH mass grows and the stellar mass becomes a smaller fraction of the total gravitational potential, the strength of 
global instabilities would diminish and angular momentum transport becomes less efficient, 
allowing continued star formation to partially deplete the inflow. 
We therefore expect more stable and long-lived phases close to the right boundary of the white region, 
towards which dynamical starburst systems tend to converge.}

{At much later times, 
the depletion of external gas fuel leads to a decline of $\dot{M}$, 
slowly quenching gravitational instability, 
and the system enters the classical quasar phase (gray region). 
Additionally, 
chemical enrichment from stellar evolution 
\citep{Cantiello2021,AliDib2023,Xu2025} can gradually introduce metallicity and dust, giving rise to FIR emission 
as we discussed in 
\S \ref{sec:universal_Teff} and will revisit in \S \ref{sec:metal_dust}. 
By this stage, 
the system evolves into classical quasars in multiple aspects 
(higher $M_\bullet$, 
standard disk and dust emission), 
completing the evolutionary cycle. }

{In this context, the angular momentum transport mechanism proposed by \citet{Thompson2005}, where}

\begin{equation}
v_r \sim m (H/R) \Omega R \gg \alpha (H/R)^2 \Omega R,
\label{eqn:globalvr}
\end{equation}

{can be interpreted as a quantitative expression of global instabilities that operate most efficiently 
in the stellar-potential-dominated regime. 
These instabilities can drive rapid, potentially sonic inflows, 
allowing the system to rapidly evolve from initial states in the red region toward more steady configurations 
in the white region. 
As the SMBH potential becomes dominant, 
their influence diminishes and they play a less significant role.}


\section{Emission From the Inner Disk}
\label{sec:inner_disk}

Most extended AGN disk models assume that the self-gravitating region terminates at an inner radius where viscous dissipation from turbulence becomes sufficient to power the local radiative output \citep{Sirko2003,Thompson2005,ChenLin2024}. 
{At this point, 
accretion heating overtakes stellar feedback, and the disk could maintain $Q>1$ solutions and quench star formation processes.}
The criterion can be defined as

\begin{equation}
\dfrac{3 G \dot{M}(R_{\rm sg}) M_\bullet}{4\pi R_{\rm sg}^3} = 2\sigma T_{\rm eff}^4(R_{\rm sg}).
\label{eqn:viscous_criterion}
\end{equation}

Within $R < R_{\rm sg}$,
the disk transitions to a standard, accretion-powered configuration.
This inner zone, 
powered solely by viscous dissipation, 
will radiate a UV-bright component with luminosity 

\begin{equation}
   L_{\rm AGN} =  3.4\times 10^{44} (\dfrac{\epsilon_\bullet}{0.06})\left(\dfrac{\dot{M}(R_{\rm sg})}{0.1 M_\odot/{\rm yr}}\right) {\rm erg/s},
\end{equation}
 
{As noted earlier, the fraction of inflow that reaches $R_{\rm sg}$, 
i.e., $\dot{M}(R_{\rm sg})$, 
is sensitive to the detailed radial structure of the disk and the assumptions adopted in solving it.
For constant $\dot{M}$, 
$L_{\rm AGN}$ from the inner disk would likely produce strongly variable UV and potentially also X-ray emission incompatible with LRDs \citep{Yue2024}, unless an additional inner truncation 
$\gg R_g$ suppresses 
the SMBH radiative efficiency $\epsilon_\bullet$ 
\citep{Liu2025}. On the other hand, with an effective conversion rate $\epsilon$ 
from accretion and formation of stellar objects \citep{Thompson2005,ChenLin2024}, }

\begin{equation}
    \dfrac{d \dot{M}}{dR}= 4 \pi  \frac{\sigma T_{\rm eff}^4}{\epsilon c^2} R \,
\end{equation}

{$\dot{M}(R_{\rm sg})$ is naturally suppressed since a converted mass fraction required to produce $L_{\rm disk}$ is}

\begin{equation}
\begin{aligned}
\frac{L_{\rm disk}}{\epsilon \dot{M} c^2} \sim\;& \dfrac{0.01}{\epsilon}\left(\frac{f}{0.2}\right) \left(\dfrac{\alpha}{0.1}\right)^{-4 / 9} \\
\times& \left(\frac{\dot{M}}{1 M_{\odot} \mathrm{yr}^{-1}}\right)^{-5 / 9}
\left(\frac{M_{\bullet}}{10^6 M_{\odot}}\right)^{2 / 3}
\left(\frac{T_{\text{eff}}}{4000 \mathrm{~K}}\right)^{5 / 2}.
\end{aligned}
\end{equation}
{This implies a critical $\epsilon_{\rm crit}\sim 0.01$ that positively correlates with $M_\bullet$ and scales down with $\dot{M}$ and $\alpha$. 
In practice, 
because most of the luminosity comes from annuli near $R_{\rm out}$, the inner disk supply depends 
sensitively on $\epsilon$. 
When $\epsilon > \epsilon_{\rm crit}$, the accretion rate remains nearly constant and the disk extends inward. 
In contrast, when $\epsilon < \epsilon_{\rm crit}$, the disk is truncated 
close to the outer boundary, 
effectively shutting off the inner disk, 
more aligned with our picture proposed in Figure \ref{fig:schematics}. 
The transition between these regimes provides 
a sharp boundary to LRD-like systems in terms of controlling the inner disk emission.}

{The boundary between smooth and truncated disks may be less abrupt if the assumption of one characteristic $\epsilon$ is relaxed. 
In fact, $\epsilon$ can vary over a wide range, 
from $\sim 10^{-4}$ - $10^{-3}$ for stellar populations \citep{Thompson2005,Tagawa2020a}, to $\sim 10^{-2}$ for massive stars, and up to $\sim 0.1$ -$1$ for stellar-mass black holes \citep{Levin2003,Gilbaum2022,Epstein-Martin2024}. 
In addition, disk/stellar winds and outflows may further reduce the effective $\epsilon$ by removing mass without contributing significant heating. 
A fixed $\epsilon$ could, in fact, lead to local and/or global instabilities that drive $\dot{M}(R)$ toward a more regulated profile, 
where different disk annuli host different characteristic $\epsilon(R)$. 
Motivated by this perspective, we provide in Appendix \ref{app:numerical} 
disk solutions from a general parameterization in which $\dot{M}$ follows a power law in $R$, 
with small-to-large decay slopes
allowing for a smooth interpolation between the large-$\epsilon$ (nearly constant $\dot{M}$) and small-$\epsilon$ (truncated disk) limits, 
and providing controlled variation of the inner accretion rate and the magnitude of associated UV luminosity. 
We emphasize, however, 
that this construction is illustrative rather than definitive, 
and the main conclusion in \S \ref{sec:outer_disk} 
is not affected by specific choice of prescriptions to solve the inner disk profile.}

{It is worth noting that the truncated-disk picture does not require accretion onto the central SMBH to vanish completely in order to remain consistent with LRD observations. Rather, the key requirement is that the residual inflow reaching small radii be sufficiently suppressed relative to standard quasars such that the associated UV/X-ray emission becomes observationally subdominant compared to that of the stellar population. 
In this regime, 
the central object is allowed to resemble a 
radiatively inefficient source analogous to the galactic center SMBH, 
whose 
high-energy emission would remain far below current observational limits if placed at the redshifts of LRDs, even if 
still weakly accreting residual gas from ambient stellar outflows \citep{Paumard2006,Yusef-Zadeh2015,Ressler2020}.}




\section{Summary and Discussion}
\label{sec:discussion}

{We have developed a framework in which the blackbody-like, dominant optical emission of LRDs arise from optically thick, dust-poor 
self-gravitating disks, 
regulated by an outer boundary $T_{\rm eff}\sim 4500$ K set by the H$^-$ opacity.
Furthermore, allowing for 
mass depletion due to star formation and feedback can suppress the inner standard disk, 
yielding spectra 
dominated by a thermalized optical bump 
while 
remaining consistent with the weak variability and lack of strong X-ray emission observed in LRDs. 
In a time-dependent picture, 
such a system hosted by a central SMBH can be interpreted as an evolutionary phase towards later classical AGNs. }

Finally, we discuss several directions for future work within this framework, 
including the role of stellar irradiation in producing the UV continuum and emission lines, 
the impact of metallicity and dust formation on the emergence of FIR-emission and transition to classical AGNs, 
cosmological abundance implications, 
and the connection to studies on stellar evolution in AGN disks. 

\subsection{UV continuum and Broad Emission Lines}

In this letter, we have focused primarily on modeling the dominant red/optical component of LRDs, but additional components in our scenario naturally explain the UV continuum emission of LRDs. 
Observationally, at least some LRDs are spatially resolved in the rest-frame UV \citep{Chen2025,Chen2025b,Baggen2025}, 
implying that at least part of the UV emission can be attributed to the host galaxy. This interpretation is also supported by spectroscopic decomposition results \citep{Sun2026arXiv260120929SLRD}.
Meanwhile, in our scenario, we expect the self-gravitating disk to give birth to a stellar population at a spatial scale comparable to the disk. 
The observed UV may come from the stars located 
in optically thin diffuse clouds that connects with the host galaxy at larger scales (see Figure \ref{fig:schematics}). 
Our scenario is distinct from the standard AGN disk-related UV emission proposed by \citet{Zhang2025}, 
and we do not expect short-term UV variability from the stellar population.  
Future work will aim to model the stellar populations in this environment in greater detail to make quantitative predictions for the UV appearance.

The origin of the broad emission lines of LRDs remains a mystery. 
In our picture, 
UV radiation from stars at or larger than the disk scale powers the broad lines. 
The stellar ionizing spectrum is expected to be softer than that of a standard accretion disk, consistent with the weakness of high-ionization lines in at least some LRDs \citep{Lambrides2024,Wang2025}. 
A stellar origin of the broad lines may also favor a lower UV-to-H$\alpha$ luminosity ratio than typical AGN, 
as recently measured for LRDs in \citet{Asada2026}. 
The virial velocity at the disk scale can be estimated by (see Equation~\ref{eqn:Rout})

\begin{gather}
\sigma = \sqrt{GM_\bullet\over R_{\rm out}} \approx 3\times10^2 \mathrm{~km}\mathrm{~s}^{-1} \times \nonumber \\ \left(\frac{M_{\bullet}}{10^6 M_{\odot}}\right)^{1 / 3}\left(\frac{\dot{M} / \alpha}{1 M_{\odot} \mathrm{yr}^{-1}}\right)^{-1 / 9}\left(\frac{T_{\text {eff }}}{4000 \mathrm{~K}}\right)^{3 / 8}\,.
\end{gather}

This will be an upper limit if the UV-emitting regions are mainly located outside the optically thick region of the disk, where $R>R_{\rm out}$. 
At low $M_\bullet$, this estimate is insufficient to account for the observed line widths, although stellar potential may introduce additional velocity dispersion.
Also, recent studies have suggested that scattering processes rather than virial motion are the main broadening mechanism (\citealt{Rusakov2025,Naidu2025,Chang2025,Torralba2025,Sneppen2026}; but see \citealt{Juodzbalis2024,Brazzini2025}), in which case the intrinsic kinematic broadening will be less important. 
The exact location and geometry of the line-producing region remain uncertain, which we regard as an important topic for future investigation.

\subsection{Metallicity and Dust Formation}
\label{sec:metal_dust}

Motivated by the non-detection of FIR emission in LRDs, 
we invoke dust-poor opacities in this study and show that they give rise to a universal outer effective temperature for optically thick, 
self-gravitating disk solutions (\S \ref{sec:universal_Teff}). 
For illustrative purposes, 
we employ zero-metallicity opacity tables in our numerical calculations, 
which leaves no dust-grain opacity below $2000$K, only a very low molecular $\rm H_2$ opacity floor. 
In this limit, 
any FIR-emitting optically thick disk solutions with $T_{\rm eff} < 2000$K \citep{Thompson2005} is avoided for reasonable midplane $\rho$ values relevant to our context (see Figure \ref{fig:Teff_thick}). 
While this represents an extreme dust-free case, the assumption can be relaxed in more general settings. 
In our framework, 
stellar evolution enriches the disk with chemicals 
and metallicity evolution should therefore be incorporated into a time-dependent picture. 
Concurrent with the growth of SMBH and evolution of stellar population, metal enrichment 
naturally leads to a decline of the abundance of LRD-like systems
towards lower redshifts and higher luminosity \citep{Inayoshi2025abundance,Ma2025,Zhuang2025}. 
Local LRD samples are found to have non-negligible metallicity and moderate mid-to-far-IR emission \citep{Lin2025},
qualitatively in agreement with this picture. 
The effect of metallicity evolution on FIR emission 
may also be generally relevant to observed nuclear color variations in JWST high-redshift quasars \citep{Li2025arXiv250512867L}.

Additional uncertainty arises from the possibility that, 
even at low temperatures, 
metals may remain predominantly in molecular form. 
In this scenario, 
the effective opacity can remain low even at non-negligible metallicities, 
subtly decoupling the broad notion of a ``dust-poor” medium from that of a narrow ``metal-poor” one. 
Exploring these possibilities and transitions in details will be an important direction for future work.

\subsection{Cosmological Abundances}
\label{sec:soltan}
An important demographic constraint on the accretion properties of LRDs comes from the classical So{\l}tan argument \citep{Soltan1982,Yu2002}, 
which relates the integrated luminosity output of growing black holes to the cumulative black-hole mass density in the local universe. 
Interpreted in this context, 
the mass density in SMBHs produced as putative LRD ``relics” $ \rho_{\rm LRD} $ is linked to the observed luminosity density of the population 
$\mathcal{L}_{\rm LRD}$ roughly by 

\begin{equation}
    \rho_{\rm LRD} \sim \dfrac{\mathcal{L}_{\rm LRD}}{c^2} \times   \dfrac{\dot{M}_{\rm acc}}{\dot{M}_{\rm lum}} \times t_{\rm LRD}
\end{equation}

where $t_{\rm LRD}$ is the characteristic cosmic time over which LRDs are abundantly observed, 
${\dot{M}_{\rm acc}}/{\dot{M}_{\rm lum}}$ denotes the ratio between the total mass accreted and the mass converted into radiation. 
Multi-band observations of LRDs have constrained their bolometric luminosities to be substantially lower than earlier estimates based on 
standard AGN bolometric corrections \citep{Greene2025}. 
With these recently revised luminosities, 
adopting a timescale of $t_{\rm LRD} \sim 10^9 {\rm years}$ corresponding to the cosmic timescale at $z\sim 5$, 
and a canonical radiative efficiency $ \epsilon_\bullet \sim 0.1$, 
the implied mass density $\rho_{\rm LRD} \lesssim 10^4 - 10^5 M_\odot $Mpc$^{-3}$ for $\mathcal{L}_{\rm LRD}\sim 10^{40} - 10^{41}$erg s$^{-1}$ Mpc$^{-3}$ 
would be below the local SMBH mass density $\rho_{\rm BH}\sim 3\times 10^5 M_\odot $Mpc$^{-3}$ estimated from other methods \citep[e.g.][]{Bernardi2010,Bernardi2013,LiepoldMa2024} and does not pose a significant So{\l}tan-type tension. 
In fact, 
it would be consistent with the interpretation that LRDs generally transition into AGNs around $z\sim 5$, a redshift 
below which their abundance undergoes a significant decline \citep{Ma2025,Umeda2025}. 

To assess whether a stellar-object-powered disk can satisfy the same demographic constraints, 
it is useful to note that the effective ratio between mass accreted and mass converted into luminosity is now

\begin{equation}
    \dfrac{\dot{M}_{\rm acc}}{\dot{M}_{\rm lum}}\sim \dfrac{1}{\epsilon},
    \label{eqn:macc_mlum_stellar}
\end{equation}

which mainly depends on the characteristic radiative efficiency of stellar objects heating the self-gravitating disk. 
At the same time, 
$\dot{M}_{\rm acc}$ should be interpreted as the total mass inflow into the nuclear region. 
If all of the generated stellar mass eventually accrete onto the central black hole, e.g. by stellar tidal disruption and engulfment associated with extreme nuclear transients \citep{Graham2025,Hinkle2025}, 
the observed local SMBH mass density would impose a lower bound 
for LRDs. 
This would disfavor disk heating dominated by a conventional stellar population \citep{Thompson2005} with $\epsilon \lesssim 10^{-3}$, 
and instead suggest that 
massive stars 
and/or stellar-mass black holes provide the dominant heating source. 
An alternative possibility is that a substantial fraction of the stellar mass formed in the self-gravitating region subsequently migrates outward or is ejected into the host galaxy, 
in which case the effective contribution to SMBH growth is reduced by a factor of $\sim  \dot{M}(R_{\rm sg})/\dot{M}(R_{\rm out})$ from Equation \ref{eqn:macc_mlum_stellar},
loosening the constraint or even quenching SMBH growth \citep{Li2025arXiv250512867L}. 
A more quantitative, 
redshift-dependent assessment of the allowed ranges of ${ \dot{M}(R_{\rm sg})}/{\dot{M}(R_{\rm out})}$ and $\epsilon$, 
as well as the properties and fate of the embedded (and potentially ejected) nuclear stellar population, 
will be explored in future studies.

\subsection{Stellar Evolution in Quasar Disks}

Finally, we comment that our model naturally appeals to more detailed studies of stellar evolution in AGN disks \citep{Cantiello2021,Dittmann2021,Wang2021,AliDib2023,ChenRenDai2023,WangJM2023,Fan2024,Liu2024,ChenLin2024,Xu2025}, a line of work that has largely developed independently of galactic-scale star formation studies. 
A self-consistent treatment of the thermal equilibrium of massive-star-heated, 
self-gravitating disks will ultimately require a new generation of radiation-hydrodynamic simulations, 
analogous in spirit to simulations of star formation and feedback on galactic scales \citep[e.g.][]{KimOstriker2017}, 
but incorporating sink-particle prescriptions 
tailored to AGN stars 
and informed by detailed numerical studies.
To capture spectral diagnostics such as line emission and Balmer features, 
radiative transfer calculations may also need to move beyond the gray approximation toward multi-group treatments that resolve the multi-frequency vertical transport of radiation from the disk midplane to the photosphere.

\section*{Acknowledgements}

Y-X. C. thanks Douglas Lin, Liang Dai, Eugene Chiang, Chung-Pei Ma, Anna de Graaff and Minghao Guo for helpful discussions. {We thank the anonymous reviewer for suggestions that significantly improved the clarity of this letter.}

\appendix

\section{Numerical solutions of disk profiles Parametrized by $\dot{M} \propto R^\gamma$}
\label{app:numerical}

{With $\dot{M}$, $M_\bullet$, $\alpha$, 
and $R_{\rm out}(\dot{M}, M_\bullet, \alpha)$ given numerical procedures 
introduced in \S \ref{sec:outer_disk}, 
we can solve a general disk profile inwards allowing $\dot{M}$ to be a general self-similar profile in the $Q\sim 1$ region 
to mimic depletion of gas by formation and growth of embedded stellar objects:}

\begin{equation}
    \dot{M}(R) = \dot{M}(R_{
    \rm out
    })(R/R_{\rm out})^{\gamma}
\end{equation}

which closes the equation set for disk profiles $\dot{M}(R), H(R), T(R), \rho(R), T_{\rm eff}(R)$. 
For demonstration, 
we provide in Appendix \ref{app:self-similar} analytical profiles for $\rho(R), T(R)$ in the radiation pressure dominated limit, which are most relevant to our results, 
as well as the corresponding $T_{\rm eff}(R)$ assuming power law opacities, 
which shows that self-gravitating disks generally exhibit flat or even inverted $T_{\rm eff}(R)$ profiles. 
As a result, the SED of such disks will be dominated by emission from $R_{\rm out}$. 

The $Q=1$ disk stops at an inner radius $R_{\rm sg}$ where viscous heating is sufficient to power the disk emission (see Equation \ref{eqn:viscous_criterion}), within which the disk connects to a standard disk with a constant $\dot{M}(R_{\rm sg})$ that extends to an inner cutoff radius of 10 Schwarzschild radii.

When $\gamma=0$, the disk profile is qualitatively similar to the constant-$\dot{M}$ solutions in \citet{Sirko2003} (although we specifically use a different metal-free opacity). 
When $\gamma>0$, 
the profile may be linked with an effective energy conversion rate at each radius as 

\begin{equation}
    \dfrac{d \dot{M}}{dR}=\gamma(\dot{M}(R_b)/R_b)(R/R_b)^{\gamma-1} =  4 \pi  \frac{\sigma T_{\rm eff}^4}{\epsilon c^2} R,
\end{equation}
where we interpret $\epsilon(R)$ as an effective, radius-dependent conversion efficiency, as elaborated in \S \ref{sec:outer_disk}.

\begin{figure*}
    \centering
 \includegraphics[width=1\textwidth]{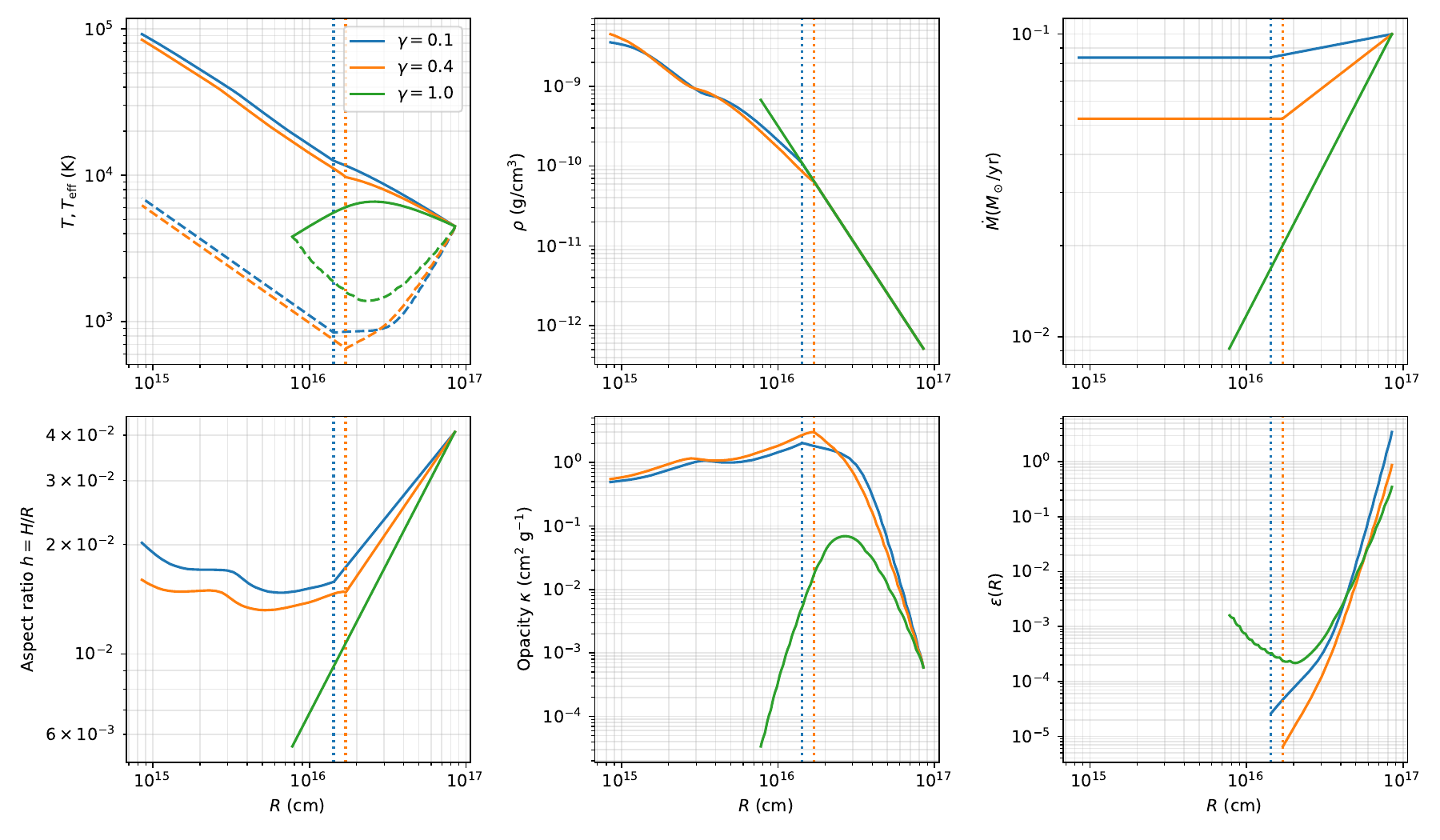}
    \caption{Radial structure of fiducial disk solutions for $M_\bullet = 10^6 M_\odot$,  $\alpha=0.1$ and outer boundary accretion rates of $\dot{M} = 0.1M_\odot$/year, with varying mass-loss slope $\gamma$. The midplane temperature $T$ and effective temperature $T_{\rm eff}$ are shown in solid and dashed lines respectively in the top left panel.
    The transition towards an inner viscous $\alpha$-disk, if present, is indicated by vertical dotted lines. } 
    \label{fig:m6a01}
\end{figure*}

In Figure \ref{fig:m6a01} we present fiducial models 
with $M_\bullet = 10^6 M_\odot$,  $\alpha=0.1$ and an outer boundary accretion rate of $\dot{M} = 0.1M_\odot$/year.
At a small decay slope $\gamma\sim 0.1$ (blue lines), 
we obtain classical solutions similar to the optically thick, 
self-gravitating region of the \citet{Sirko2003} disk model where
mass removal is modest, and  
the local transport equation yields $H\propto R^{3/2}$ and $c_s = {\rm Const}$. 
For radiation pressure dominated regions, 
the midplane temperature decreases with $R$  (or increases inwards), 
and the disk extends to $R_{\rm sg} \sim 0.2 R_{\rm out}$ (indicated by vertical dotted lines) before it transitions into an inner $\alpha$-disk. 
Between $R_{\rm sg}$ and $R_{\rm out}$, 
we expect the disk emission to be dominated by the effective temperature close to the outer boundary due to an increasing $T_{\rm eff}(R)$, 
with cooler annuli contributing to
a long-wavelength tail in the SED. 
With an increasing decay slope $\gamma$, 
the outer radius and total luminosity of the outer disk remain largely unchanged, 
as predicted in \S~\ref{sec:outer_disk}. 
However, 
a larger fraction of the inflowing mass is lost or converted into stars, making the inner disk progressively less luminous and more consistent with LRD-like properties.


Eventually at sufficiently large $\gamma$, a qualitative transition occurs: 
the disk reaches an optically thin branch before the self-gravitating radius, 
at a radius $R_{\rm thin}>R_{\rm sg}$. This behavior is evident in the $\gamma = 1.0$ case for $\dot{M} = 0.1 M_\odot$/yr in Figure \ref{fig:m6a01} (green opaque lines). 
This is due to the strong depletion of $\dot{M}$ that yields $\partial T(R)/\partial R>0$ 
within a certain radius 
and makes $T_{\rm eff}$ and $T$ converge inwards.
When this happens, we terminate the iterative solution at $R_{\rm thin}$. 
We do not attempt to 
model the detailed physics in this optically thin region which is uncertain, 
just as we do not model the optically thin cloud exterior to $R_{\rm out}$.

Regardless of what additional star formation might occur inside $R_{\rm thin}$, 
any remaining accretion that does reach the viscous inner disk would produce, at most, an AGN luminosity of 

\begin{equation}
    L_{\rm AGN, max} = \epsilon_\bullet \dot{M}(R_{\rm thin}) c^2
\label{eqn:L_AGNmax}
\end{equation}

where the actual AGN luminosity can be between 0 and $L_{\rm AGN, max}$, 
depending on how much accretion rate is converted to stars within the region $< R_{\rm thin}$. 
In practice, 
for radiation pressure dominated disks, in the presence of $R_{\rm thin}$ which only occurs at high $\gamma$, the luminosity of the outer, optically thick disk or 
``ring"
will typically exceed that of the inner components even if the residual accretion were fully converted into standard disk UV emission, and the disk is effectively truncated.

{In all cases, the implied $\epsilon(R)$ spans a wide range in the self-gravitating region (lower right panel of Figure \ref{fig:m6a01}), 
implying that global regulation processes need to be at work to maintain a self-similar $\dot{M}$ profile, 
rather than the burning of one particular population of stars. 
This contrast may be quite extreme, but since it's realized over a relatively limited radial extent (a factor of $\sim$3 - 5 for $\epsilon$ to drop from $\sim 1$ to $\sim 10^{-4}$), 
rather than across many decades, 
efficient communication between adjacent radii with different characteristic $\epsilon$ 
is not impossible. 
Moreover, in practice, because most star formation and mass depletion occur near the outer boundary, 
imposing a lower bound on $\epsilon$ 
(e.g., $\epsilon > 10^{-5}, 10^{-4}$ or $10^{-3}$) to 
limit its dynamic range does not significantly affect the disk profiles. 
We therefore view these solutions as illustrative, designed to 
interpolate between the limiting cases of constant $\dot{M}$ (large $\epsilon$) 
and strongly truncated (small $\epsilon$) disks, 
rather than as definitive physical models. }

\begin{figure*}
    \centering
 \includegraphics[width=1\textwidth]{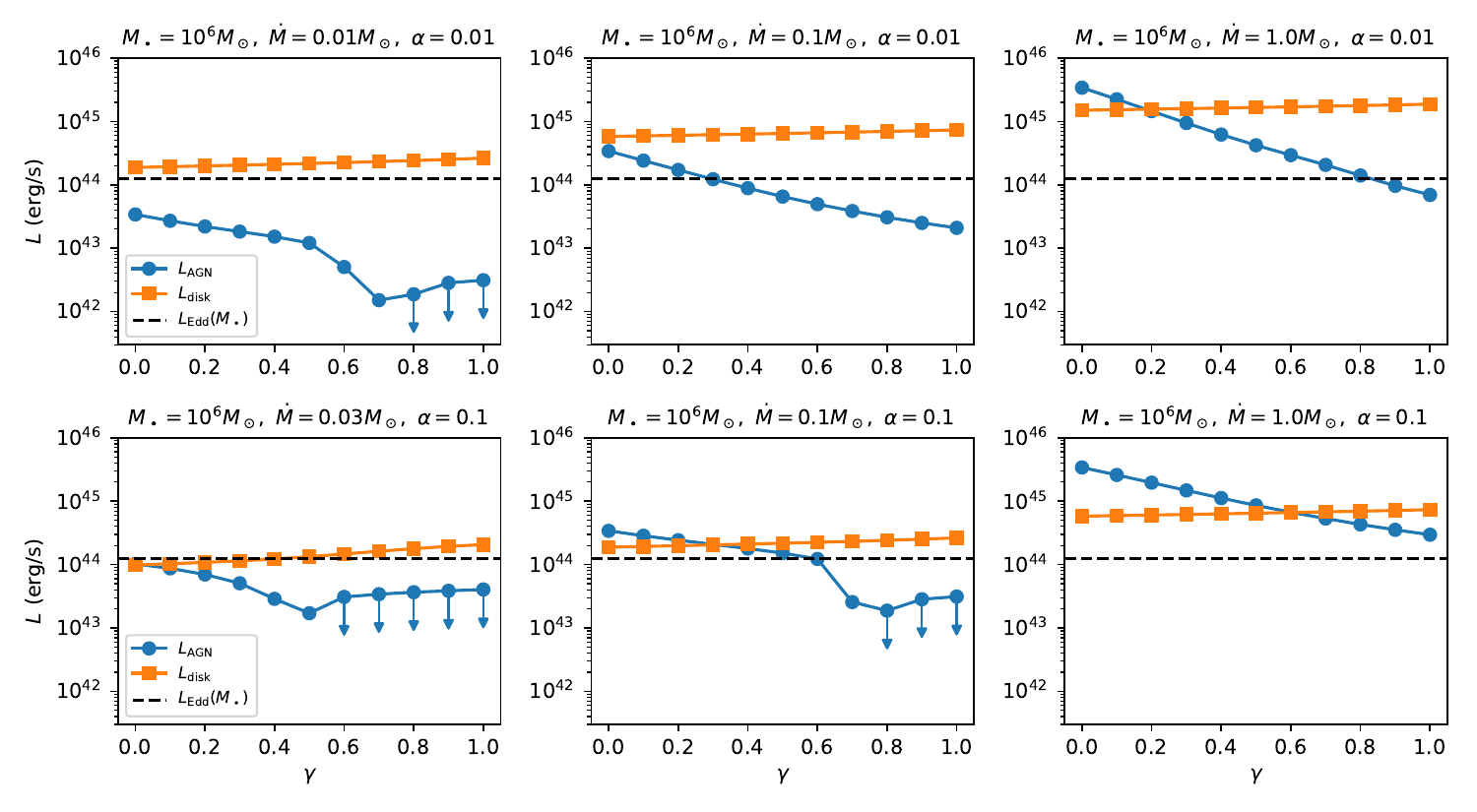}
    \caption{Summary of luminosity contributions for the accretion disk around a $10^6 M_\odot$ SMBH. The AGN component or its upper limit is shown in blue and the thermal emission from the optically thick self-gravitating region is shown in orange. The Eddington luminosity is plotted for reference (black dashed).
    Across all models, increasing $\gamma$ systematically suppresses $L_{\rm AGN}$, while $L_{\rm disk}$
    remains largely unchanged and often dominates even at $\gamma$ for lower accretion rates. } 
    \label{fig:m6_Lsummary}
\end{figure*}

To more quantitatively demonstrate the decrease of the inner disk's emission 
with  $\gamma$, 
in Figure \ref{fig:m6_Lsummary} we plot $L_{\rm AGN}$ or $L_{\rm AGN, max}$ (as upper limits at sufficiently high $\gamma$), 
as well as the thermal emission of the optically thick self-gravitating disk region $L_{\rm disk}$ 
for comparison, 
defined to exclude contributions from the standard inner disk and any optically thin zones
as a function of $\gamma$ for a variety of $\dot{M}$ and $\alpha$, including the two cases shown in Figure \ref{fig:m6a01}. 
The trend clearly demonstrates that in all cases, as $\gamma$
increases, 
the standard-AGN luminosity declines while the outer-disk emission dominates. 

\begin{figure*}
    \centering
 \includegraphics[width=1\textwidth]{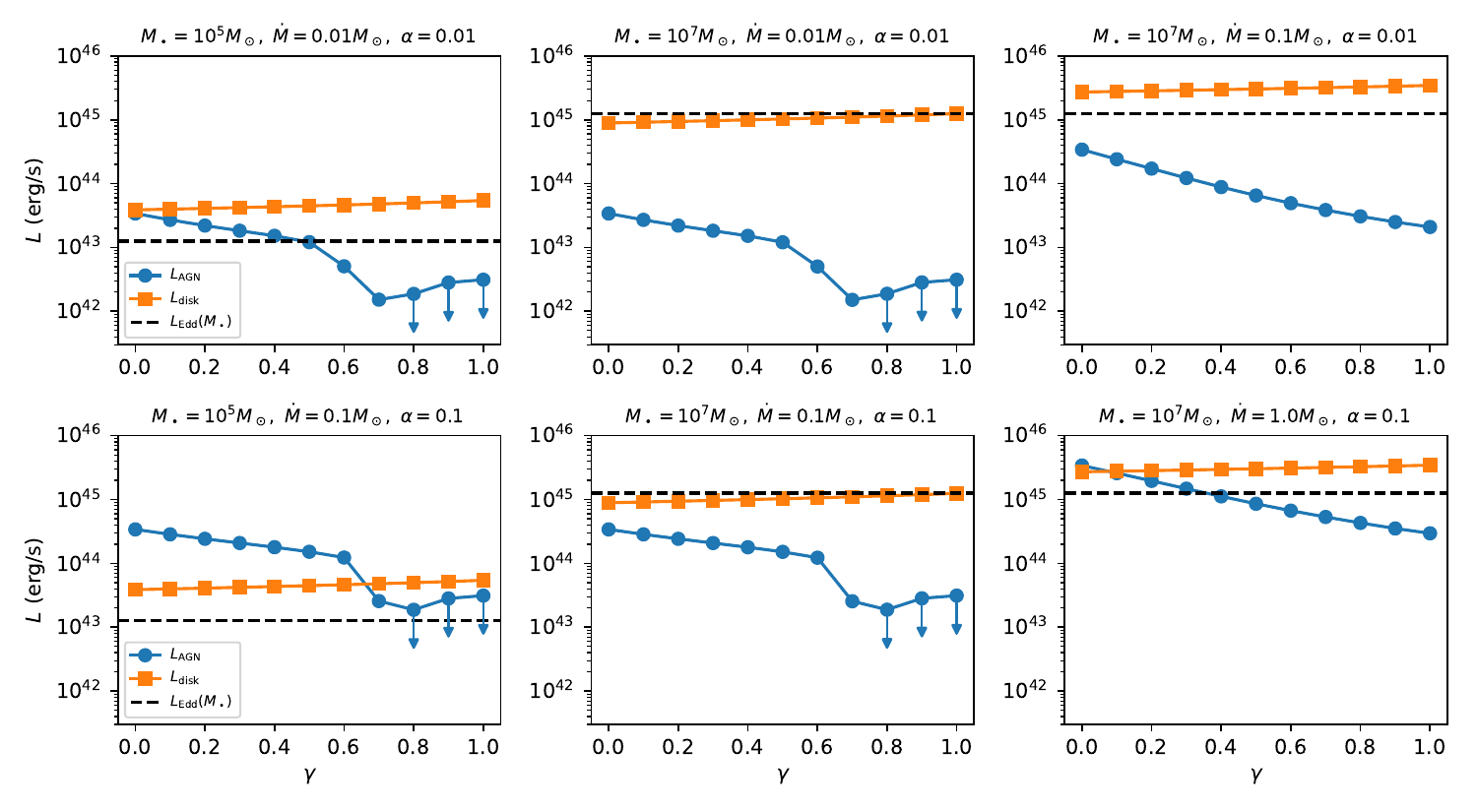}
    \caption{Summary of luminosity contributions for the accretion disk around a $10^7 M_\odot$ or $10^5 M_\odot$ SMBH for certain accretion rates and $\alpha$. } 
    \label{fig:m75_Lsummary}
\end{figure*}

{Orthogonal to the constraint on $\gamma$ set by $L_{\rm AGN}$, the outer-disk luminosity $L_{\rm disk}$ depends primarily on $\dot{M}/\alpha$, in agreement with the scalings derived in \S~\ref{sec:outer_disk}. Along the SMBH mass dimension, Figure~\ref{fig:m75_Lsummary} compares the luminosity components for several choices of $\dot{M}$ and $\alpha$, 
varying the SMBH mass between $M_\bullet = 10^7,M_\odot$ and $10^5,M_\odot$. 
At fixed $\dot{M}/\alpha$, the ratio $L_{\rm disk}/L_{\rm Edd}$ decreases with increasing SMBH mass, consistent with Equation~\ref{eqn:Ldiskanalytical}. Conversely, systems with lower $M_\bullet$ are more likely to enter the stellar-potential-dominated regime (Equation~\ref{eqn:Maupperlimit}).}

\begin{figure*}
    \centering
 \includegraphics[width=1\textwidth]{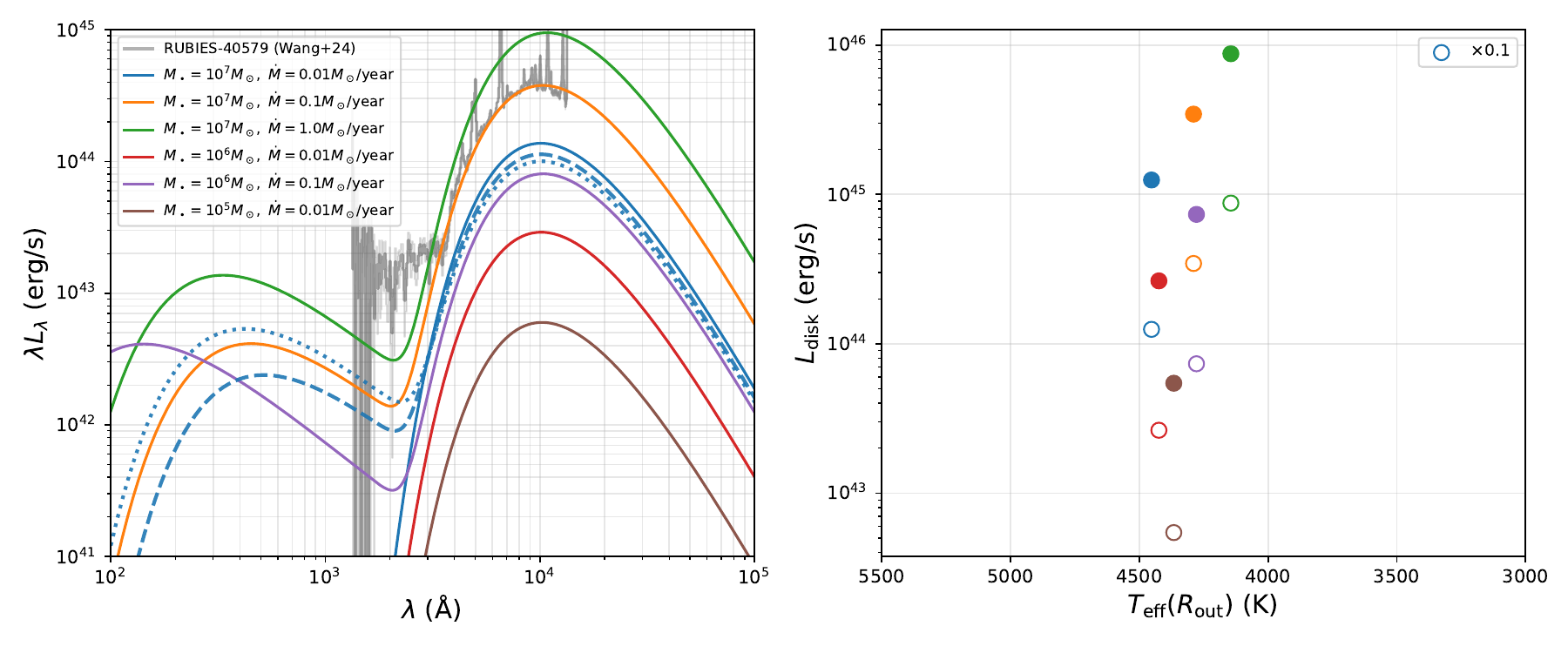}
    \caption{Left: Example SEDs for selected $\alpha=0.01$ disk models by integrating blackbody emission over each disk annuli. All solid lines assume $\gamma=1$ while the dashed and dotted lines correspond to $M_\bullet = 10^7M_\odot, \dot{M} = 0.01 M_\odot$/yr model with $\gamma = 0.5, 0.1$ respectively which allows for more UV radiation from the inner AGN disk. 
    We vary $\dot{M}, M_\bullet$ and not $\alpha$ since we expect from Equation \ref{eqn:Ldiskanalytical} that disks with similar $\dot{M}/\alpha$ 
    at given $M_\bullet$ will have similar emission and so they can be scaled to higher viable $\alpha$ values easily. A representative LRD spectrum (RUBIES-UDS 40579) is plotted in gray. 
    Right: disk luminosity versus $T_{\rm eff}(R_{\rm out})$ for
    different $\dot{M}, M_\bullet$ parameters, 
    with hollow symbols being scaled by 0.1x to naively demonstrate effects of the viewing angle. The self-gravitating disk naturally shows a ``Hayashi limit''. 
    } 
    \label{fig:SED}
\end{figure*}

All our optically thick disk models naturally yield characteristic $T_{\rm eff}\approx 4500$K at their outer truncation radii.
This temperature regulation ensures that the disks robustly produce a red optical component similar to LRDs 
without requiring additional parameter tuning.
To emphasize this point, 
we compute the full SEDs by integrating the thermal emission over all disk annuli from $R_{\rm out}$ to i) 10 Schwarzschild radii of the SMBH if a standard $\alpha$-disk is present or ii) to $R_{\rm thin}$ if otherwise. 
The observed spectrum of a representative LRD, RUBIES-40579 \citep{WangdeGraaff2025}, 
is shown in the background for comparison, illustrating the similarity between the predicted red/optical bump and the observed LRD SED.
The left panel of Figure \ref{fig:SED} presents the SEDs for representative disk models spanning a range of ($M_\bullet$, $\dot{M}$) at fixed $\alpha = 0.01$. The solid curves correspond to models with $\gamma = 1$ and demonstrate the universal red/optical bump across different parameter choices. 
The dashed and dotted curves have $\gamma = 0.5$ and $\gamma = 0.1$, respectively, for a given parameter $M_\bullet = 10^7M_\odot, \dot{M} = 0.01 M_\odot$/yr, illustrating how the inner disk's UV emission progressively weakens as $\gamma$ increases. 
The right panel shows the corresponding outer disk luminosities, $L_{\rm disk}$, 
plotted against the effective temperature at the outer radius, 
which sets the peak of the red/optical emission and illustrates the emergence of a ``Hayashi-limit" for self-gravitating disks across 2-3 order of magnitudes in luminosity. 
Hollow circles indicate a factor of $\sim 0.1$ reduction in the inferred luminosity, 
representative of plausible viewing angle effects due to $H/R\sim 0.1$. 
We also comment that such optical/red bumps, 
if fitted as a modified blackbody $F_\nu \propto B_\nu(T) \nu^{\beta}$, 
will most likely prefer a broader $\beta\leq0$ due to the initial decay of $T_{\rm eff}(R)$ inwards down to $1000-2000$K, 
consistent with a considerable number of samples in \citet{deGraaff2025}. 
This is also consistent with LRD samples presented in \citet{Wang2026}, 
whose emission contain contribution from $T_{\rm eff} \lesssim 3000$K atmosphere components with water absorption features. 
Nevertheless, 
we caution that the exact shape of the thermal emission depends on radiative transfer effects and may deviate from a blackbody \citep{Liu2026}, 
and detailed fitting will not be the focus of this letter.

\section{Analytical $\rho, T, T_{\rm eff}$ solutions for a radiation pressure dominated, self-gravitating disk}
\label{app:self-similar}
Given the disk equations in \S \ref{sec:universal_Teff} and \S \ref{app:numerical}, 
the density and temperature profiles in a radiation pressure dominated $Q=1$ disk with power law accretion rate profile $\dot{M} \propto R^\gamma$ has the following power law solution

\begin{equation}
    \rho = \rho_0 (R/R_{\rm out})^{-3}, T = T_0 (R/R_{\rm out})^{-3/4 + \gamma/6}
\end{equation}

where the normalizations are 

\begin{equation}
   \rho_0 = \dfrac{M_\bullet}{2\pi  R_{\rm out}^3}, T_0 = \left[\dfrac{G \dot{M}_{\rm out}}{\alpha}(\dfrac3{2\pi a})^{3/2}\right]^{1/6} M_\bullet^{1/4} R_{\rm out}^{-3/4}
\end{equation}

setting $\kappa = \kappa(\rho_0, T_0) (\rho/\rho_0)^p (T/T_0)^q$, 

\begin{equation}
    T_{\rm eff} = T_{\rm eff,0} (R/R_{\rm out})^{\eta}, \eta = (-\dfrac3{16}+\dfrac\gamma{24})(2-q) + \dfrac{3p}4 
\end{equation}

where

\begin{equation}
    T_{\mathrm{eff},0} = \left[\dfrac{T_0^2}{\kappa_R(\rho_0, T_0) \sqrt{\frac{a}{6 \pi G }}}\right]^{1 / 4}
\end{equation}

In summary, the effective temperature profile of a self-gravitating disk is insensitive to the detailed accretion-rate radial dependence encoded by $\gamma$, 
which enters only weakly into the power-law index of $T_{\rm eff}(R)$. 
For constant opacity ($p=q=0$), we obtain $\partial \ln T_{\rm eff}/\partial \ln R \simeq -3/8$ \citep{Sirko2003}, which is already significantly flatter than the canonical $-3/4$ scaling of a standard thin disk. We note that $\partial \ln T_{\rm eff}/\partial \ln R>-1/2$ implies that the outer edge dominates the total radiative output.
Adopting a Kramers opacity ($p=1$, $q=-3.5$) further flattens the profile to $\partial \ln T_{\rm eff}/\partial \ln R \gtrsim -0.3$, albeit still declining. 
In contrast, when the opacity is dominated by H$^{-}$ with $q \sim 9$, 
the resulting slope becomes positive, leading to an inverted effective temperature profile. 
In all cases, these scalings imply that self-gravitating disks generically have SEDs dominated by emission from their outer radii. 

\bibliography{sample631}{}
\bibliographystyle{aasjournal}



\end{CJK*}
\end{document}